\documentclass[osajnl,longbibliography,twocolumn,showpacs,superscriptaddress,10pt]{revtex4-1} 
\usepackage{amsmath,amssymb,graphicx}
\usepackage[final]{hyperref}
\hypersetup{
	colorlinks=true,       
	linkcolor=blue,          
	citecolor=blue,        
	filecolor=magenta,      
	urlcolor=blue         
}
\begin{document}

\title{Generating squeezed vacuum field with non-zero orbital angular
momentum with atomic ensembles}

\author{Mi Zhang}
\author{Joseph Soultanis}
\author{Irina Novikova}
\author{Eugeniy E. Mikhailov}
\affiliation{The College of William $\&$ Mary, Williamsburg VA 23187 USA}

\begin{abstract}

We demonstrated that by using a pump field with non-zero orbital
angular momentum (OAM) in the polarization self-rotation squeezing process  it is possible to generate a squeezed vacuum  optical field with the matching OAM. 
We found a similar level of maximum quantum noise reduction for 
a first-order Laguerre-Gaussian pump beam and a
regular Gaussian pump beam, even though the optimal operational conditions
differed in these two cases. 
Also, we investigated the effect of self-defocusing
on the level of the vacuum squeezing by simultaneously monitoring the minimum
quantum noise level and the output beam transverse profile at various pump laser
powers and atomic densities, and found no direct correlations between
the increased beam size and the degree of measured squeezing.  
\end{abstract}


\maketitle


Light beams which carry an orbital angular momentum (OAM) have recently gained
popularity for many optical
applications~\cite{OVbook,StructuredLightBook,Torres2007NaturePhy}.  For example, OAM provides an additional degree of freedom for an optical
field (in addition to traditional frequency and polarization), that can be
used to increase the information capacity of an optical
network~\cite{TBvortexNatPhot2012}.  It also allows generation of
entanglement between a pair of single
photons~\cite{ZeilingerNature01,tornerOL04,BarbosaPhysRevLett05,Torres2007NaturePhy}
or between continuous optical
fields~\cite{LettPRL08,MarinoPhysRevLett08,BachorNaturePhotonics09} for
spatial multimode quantum information systems and imaging. Moreover, a
hyperentanglement  between spin and orbital angular momentum states of a
photon~\cite{KwiatNP08,ChenOL09} has been demonstrated to increase the
dimensionality and capacity of quantum channels.    

Here we demonstrate a simple way to generate an optical squeezed
vacuum  field with a non-zero OAM via interaction of a
linearly polarized Laguerre-Gaussian pump field with a resonant atomic
vapor under the polarization self-rotation (PSR) conditions~\cite{matsko_vacuum_2002,ries_experimental_2003,mikhailov2008ol}.
Previous experiments in PSR squeezing have demonstrated quadrature
noise suppression in the vacuum field in the orthogonal polarization up to
3~dB below the shot-noise
limit~\cite{lezama2011pra,mikhailov2012sq_magnetometer}. 
In our experiments we used a spiral phase mask to convert
the pump field into a first-order Laguerre-Gaussian beam
before the interaction with Rb atoms, then analyzed the quantum noise in the
orthogonal polarization after the vapor cell using the same pump field as a
local oscillator. In this case we detected up to  $1.7\pm 0.2$~dB of quantum noise
suppression in the matching spatial mode. This value is comparable to the $1.8\pm
0.2$~dB of squeezing measured in the same vapor cell using a pump
field with a regular Gaussian distribution.
It is worth mentioning that a similar strategy of using an OAM pump beam
was used previously for the generation of photon pairs with OAM via parametric
down conversion~\cite{ZeilingerNature01,BarbosaPhysRevLett05}, and more
recently in demonstration of intensity-squeezed bright twin beams with
non-zero OAM~\cite{MarinoPhysRevLett08} via a non-degenerate four-wave
mixing process.

 \begin{figure}[h]
	\includegraphics[width=1.0\columnwidth]{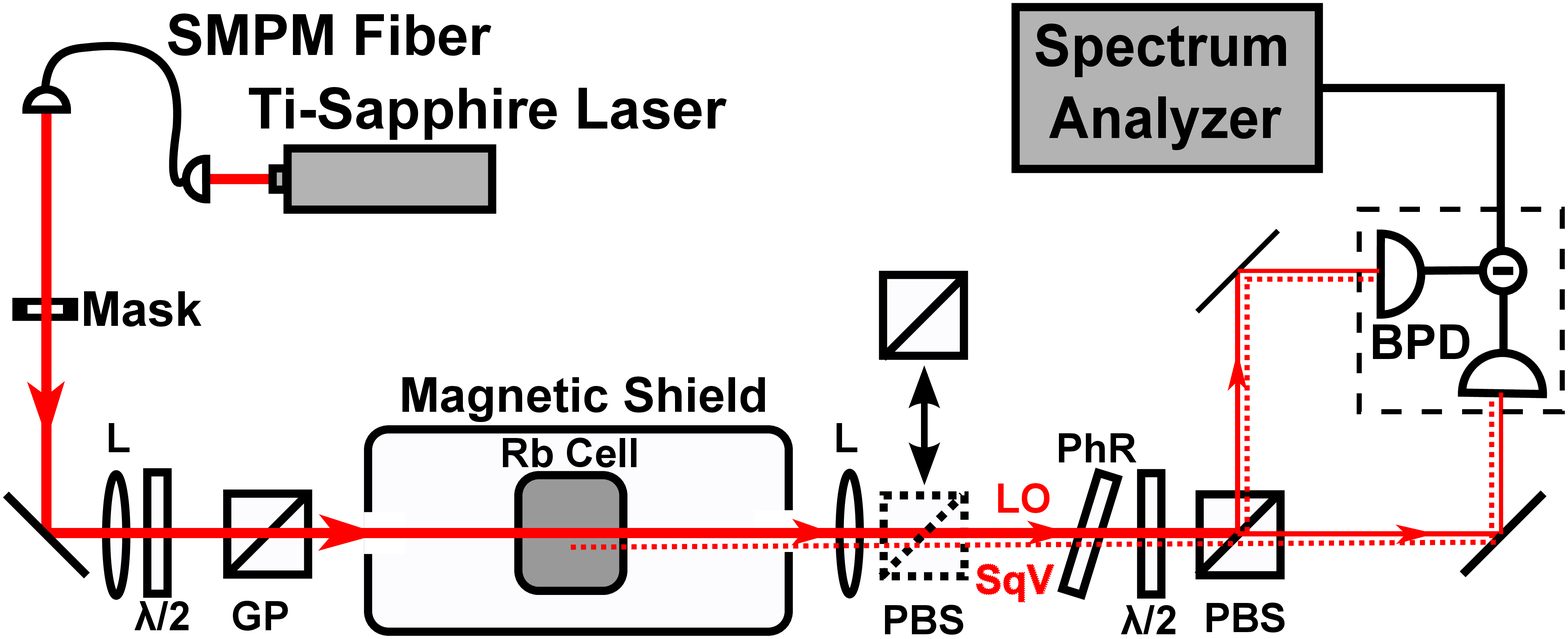}
	\caption{
		\label{fig:setup}
		Experimental setup. 
		SMPM fiber depicts single-mode polarization-maintaining fiber,
		$\lambda/2$ is half-wave plate,
		GP is Glan-laser polarizer,
		PBS is polarizing beam splitter,
		PhR is phase-retarding wave plate,
		and BPD is balanced photodetector.
	}
\end{figure}

The schematic of the experiment is shown in Fig.~\ref{fig:setup}. 
The output of a cw Ti:Sapphire laser was tuned near the $5^{2}S_{1/2} F=2
\rightarrow 5^{2}P_{1/2},F^{\prime} = 2$ transition of the ${}^{87}$Rb
($\lambda \simeq$ 795~nm). 
We used a single-mode optical fiber followed by a Glan-laser polarizer (GP)  to
prepare a high quality linearly polarized pump beam with  the Gaussian transverse profile, which then was focussed inside a
cylindrical Pyrex cell (10~mm in length and 25~mm in diameter) containing isotopically enriched $^{87}$Rb vapor. 
The focal lengths of the
lenses before and after the cell were correspondingly $40$~cm and $50$~cm. 
The size of the minimum focal spot inside the cell was $0.13\pm 0.01$~mm FWHM.  
The vapor cell was mounted inside a three-layer magnetic
shielding, and the number density of Rb atoms was adjusted between
$3.4\cdot 10^{11}~\mathrm{cm}^{-3}$ and $6.0\cdot
10^{12}~\mathrm{cm}^{-3}$ by adjusting the cell's temperature. 
The input laser
power in the cell was controlled by rotating a half wave plate before
the Glan polarizer, with maximum injection power $16$~mW.

We analyzed the quantum noise of the vacuum field in
orthogonal linear polarization (with respect to the pump field) after
the Rb cell by means of a homodyne
detection~\cite{lezama2011pra,mikhailov2012sq_magnetometer}. 
We reused the strong pump field as the local oscillator (LO), avoiding
spatial separation of the LO and the vacuum optical field (SqV) to improve the
stability of the detection. 
To achieve this we rotated the polarizations of both optical fields
by $45^\circ$ with respect to the axes of a polarizing beam splitter (PBS). The
relative phase between the two polarizations was adjusted to detect
minimum noise quadrature by horizontally tilting  a phase-retarding plate (PhR) -- a
quarter-wave plate with optical axes aligned with the local oscillator and
the vacuum field polarizations.  The two outputs were then directed to a
balanced photodetector (BPD) with $1.6\times 10^{4}$~V/A gain, 9~MHz 3~dB
bandwidth, and dark noise level at least 10~dB below the shot noise level. 
The shot noise level measurements were done with a polarizing beam
splitter placed after the Rb cell such that only the pump field was
transmitted, and the modified vacuum field in the orthogonal polarization
was rejected.  

\begin{figure}[h]
    \includegraphics[width=1.0\columnwidth]{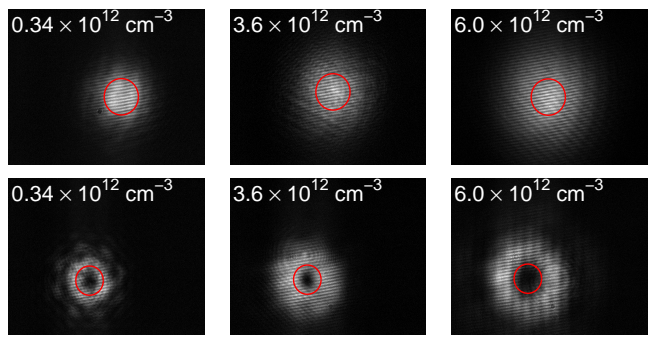}
    \caption{
             \label{fig:gallery}
						 The transverse profiles of a Gaussian (top) and vortex (bottom) beams
						 after interaction with the Rb vapor cell at different atomic
						 densities.
						 The red (light grey) circles are shown to aid visual
						 comparison of beam sizes in low and high atomic density
						 cases for the Gaussian and vortex beams, correspondingly.
             }
            
\end{figure}

To modify the transverse profile of the pump beam and add a
non-zero OAM, we placed a spiral phase mask in the collimated
portion of the beam path before the Rb cell, as shown in
Fig.~\ref{fig:setup}. 
The azimuthal thickness variation of the mask produced a $2 \pi$
phase difference, creating a phase singularity at the center of the
transmitted laser beam. As a result, its radial intensity distribution
dropped to zero at the center (so called ``optical vortex'')~\cite{OVbook},
forming a signature ``donut''-shaped transverse profile shown in
Fig.~\ref{fig:gallery}. These images were recorded by a CCD camera placed
after the Rb vapor cell. 
In general, the recorded intensity distributions were well described by
the first order Laguerre-Gaussian distribution, characteristic for the
laser beam carrying $1 \hbar$ angular momentum:
\begin{equation}\label{LGdistribution}
I(r) = I_{0}\frac{2r^{2}}{w^{2}} e^{-\frac{2r^{2}}{w^{2}}},
\end{equation}
where $w$ is the waist of the vortex beam, and $\pi w^2 I_{0}/2$ is the
total power.  The variation in the mask's thickness was not smooth, but
changed step-like through 8 discreet sectors, causing small additional
features outside of the main vortex beam due to the diffraction of
light on the boundaries of the phase mask sectors. 
Without the mask, the transverse intensity profile of the laser beam is
accurately described by the regular Gaussian distribution:  
\begin{equation}\label{Gdistribution}
I(r) = I_{0}e^{-\frac{2r^{2}}{w^{2}}}.
\end{equation}

\begin{figure}[h]
    \includegraphics[width=1.0\columnwidth]{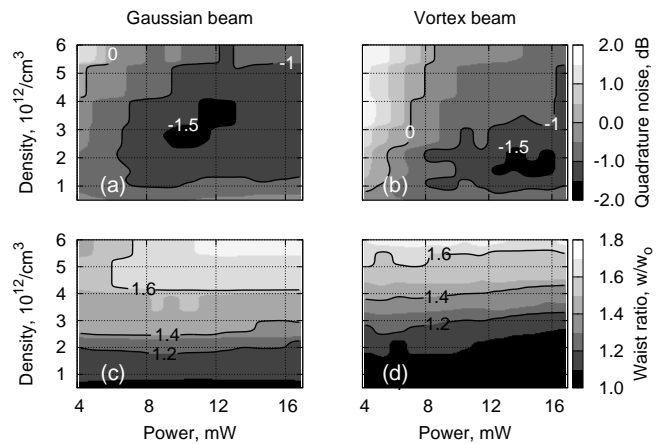}
   \caption{
			\label{fig:sqz_and_waist_vs_temp_and_power}
			Measured minimum quadrature noise power (top row) and the relative
			beam expansion (bottom row) for the pump beam with the Gaussian
			(left column) and Laguerre-Gaussian (right column) distributions as
			functions of the pump power and the atomic density. 
			The beam expansion was measured as  the ratio of the measured waist
			($w$)
			[from fits (\ref{LGdistribution}) and (\ref{Gdistribution})] to its
			value at low temperature ($w_0$), where self-defocusing was negligible.
			For quantum noise measurements, zero corresponds to the shot noise
			level.  Spectrum analyzer detection frequency was $1$~MHz.
	}
\end{figure}

Previous experiments show that PSR-based squeezing requires careful
optimization of the experimental parameters, such as atomic density,
laser frequency, power and focusing characteristics inside the vapor
cell~\cite{mikhailov2009jmo,horrom2013thesis}, these optimal
conditions change depending on the geometry and the buffer gas composition
of a Rb vapor cell.  
To identify these optimal conditions in the current experimental setup, we
mapped the dependence of the minimum measured quantum noise power as a
function of the laser power and the atomic density. For each measurement we optimized the laser frequency for the highest value of squeezing, withing approximately $200$~MHz around the center of the atomic resonance. 
The results of these measurements are shown in
Fig.~\ref{fig:sqz_and_waist_vs_temp_and_power}(a,b).  
For a regular pump beam with a Gaussian transverse distribution
[Fig.~\ref{fig:sqz_and_waist_vs_temp_and_power}(a)], the best recorded
squeezing of $1.8\pm 0.2$~dB was observed at a pump power of 10.5~mW
and the atomic density of a $2.7\times 10^{12}\mathrm{~cm}^{-3}$. 
The measured squeezing level was somewhat worse than previously observed
values at this Rb optical transition~\cite{mikhailov2012sq_magnetometer}; possibly due to higher cell temperature (to compensate for shorter cell length). 
Similar to the previous observation, the maximum squeezing occurred is a small ``island'' of the  
pump power/ atomic density parameter space.

We then repeated the same procedure using a Laguerre-Gaussian pump beam.
Fig.~\ref{fig:sqz_and_waist_vs_temp_and_power}(b) shows the minimum
quadrature noise power at different values of the laser power and atomic
density. The minimum quantum noise level, detected with the optical
vortex pump beam was $1.7\pm 0.2$~dB below the shot noise.  Since the same OAM pump beam 
was used as the LO in the homodyne detection,  we conclude
that the squeezed vacuum optical field also was carrying the same OAM $1\hbar$.
This observation is consistent with the conservation of the angular momentum.  Previous
experiments have demonstrated that the OAM is conserved in  four-wave
mixing processes~\cite{lettPRA08,MarinoPhysRevLett08}.  The generation of the PSR squeezing
can be described as a degenerate four-wave
mixing~\cite{mikhailov2009jmo}, in which two photons of one linear
polarization are absorbed from the pump field, and a pair of photons is
emitted in the correlated noise sidebands of the orthogonal polarization.
As each of the four photons involved in the process can carry the same
angular momentum $1\hbar$, the total angular momentum is conserved. 

\begin{figure}[h]
	\includegraphics[width=1.0\columnwidth]{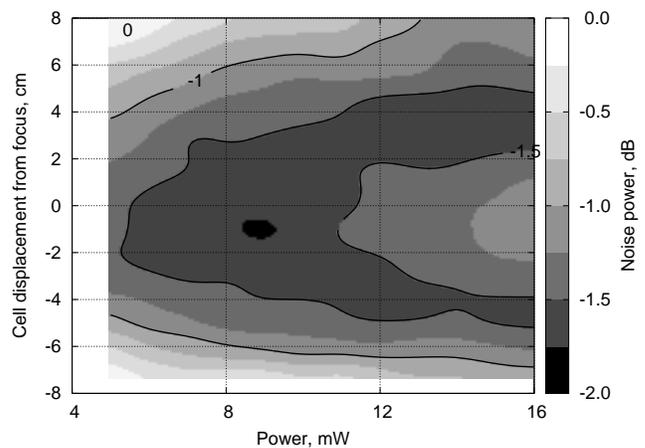}
   \caption{
			\label{fig:cell_shift}
			Measured minimum quadrature noise power as a function of position of
			the focal point and the pump intensity. 
			Zero displacement corresponds to the Gaussian pump laser focused at the
			center of the vapor cell.  
			For quantum noise measurements zero corresponds to the shot noise
			level.  Spectrum analyzer detection frequency was $1$~MHz.
	}
\end{figure}

The optimized value of measured squeezing with OAM pump beam matched the value obtained using a regular pump beam within the
experimental uncertainty.
At the same time, the optimal experimental conditions differed in these two cases.
For the vortex pump beam the best squeezing of $1.7\pm 0.2$~dB occurred at a higher
optical pump power of $14.7$~mW and a lower atomic density of
$(1.8 \pm 0.3)\times 10^{12}~\mathrm{cm}^{-3}$.  (Under identical conditions, the
squeezing obtained with a regular pump beam was only  $1.1\pm 0.2$~dB.)
Such changes in optimal experimental parameters was not surprising,
since the details of the pump beam propagation inside the atomic ensemble
were known to have a strong effect on the output squeezed vacuum.  For
example, Fig. ~\ref{fig:cell_shift} shows the variations in the
measured squeezing as the magnetic shield, containing the vapor cell,  was
shifted back and forth along the focused Gaussian pump beam path.
Considering the depth of focus of approximately $4.8$~cm, it is easy to
see that the best value of squeezing was obtained with the lowest pump
power when the cell was positioned around the focal point.  Any
displacement of the cell away from the focus in either direction resulted
in achieving similar value of squeezing at higher value of the pump power.
Since the peak intensity of the first-order Laguerre-Gaussian beam is less than half of the peak intensity of a regular Gaussian beam with
the same waist parameter, we expect to see a higher laser power to produce
optimal squeezing for the vortex pump beam.

Our experimental arrangement also allowed us to investigate the effect of
self-defocusing of the optical beams in Rb vapor at higher atomic density.
Self-defocusing/self-focusing is a well-known nonlinear
effect~\cite{Grischkowsky1972pra,AshkinPhysRevLett74} when a strong optical
field propagating through a resonance optical medium induces an
intensity-dependent variation in its refraction index; thus, a transverse
intensity distribution of an optical field, ``mapped'' into a spatial
variation of the refraction index, creates an effective atomic lens that
changes the size and divergence of the output optical beam.
Fig.~\ref{fig:gallery} clearly shows that we observed a strong defocusing
effect for both regular and vortex pump beams, which was more pronounced at higher
densities of Rb atoms.  Previous work showed (both
experimentally and theoretically) that such beam distortion can limit the generation of squeezed vacuum  in
the four-wave-mixing process~\cite{shapiro91ol}. 

To search for correlations between the beam size variation and observed
squeezing level, we recorded the images of the output pump beam intensity
distributions for different values of laser power and atomic density
matching the experimental parameters of the squeezing level measurements
depicted in Fig.~\ref{fig:sqz_and_waist_vs_temp_and_power}(a,b).
Since the intensity
distributions of all beams were well-fitted by either
Eq.(\ref{LGdistribution}) (with phase mask inserted) or
Eq.(\ref{Gdistribution})(with no phase mask), the measurements of the
waist parameter $w$ were sufficient to accurately describe beam
modifications at various experimental parameters. The results of these
measurements are shown in
Fig.~\ref{fig:sqz_and_waist_vs_temp_and_power}(c,d) for both Gaussian and
Laguerre-Gaussian pump beams. 

In our detection scheme we used the output pump field as a local
oscillator, substantially reducing the sensitivity to the beam distortions
(compared to an independent LO beam in Ref.~\cite{shapiro91ol}) as long as
both the squeezed vacuum and the pump field were spatially mode-matched.   A
simple comparison of the data in
Figs.~\ref{fig:sqz_and_waist_vs_temp_and_power}(a) and (c) reveals that
the observed maximum squeezing occurred at the region of moderate
($\approx 50~\%$) beam expansion for the Gaussian beam. The same is true for
the OAM pump beam [Figs.~\ref{fig:sqz_and_waist_vs_temp_and_power}(b) and
(d)].  For a fixed atomic density there is very little variation in the
beam diameter with respect to the laser power.  Simultaneously, the measured values
of squeezing showed much stronger intensity dependence, with squeezing
reaching a local maximum at some intermediate power, and then decreasing
at higher powers.  These observations somewhat contradict the detailed
theoretical calculations~\cite{mikhailov2009jmo} that the value of
squeezing must continuously grow with laser power.  At the same time, it
cannot be explained by the self-defocusing effect either, since the size
of the laser beam does not change at the higher intensities compared to
the optimal intensity at fixed atomic density. 
Thus, based solely on these measurements we cannot
completely rule out the self-focusing effect, since both beam
expansion and squeezing deterioration become more pronounced at high
atomic densities.  It is possible that as atomic density increases, the
spatial modes for squeezed vacuum and the pump field may experience
different defocusing, resulting in the reduction in the measured squeezing
due to the mode-mismatch at the detection stage.  To unambiguously
distinguish such differential self-defocusing effect from other nonlinear
interactions, such as spontaneous Raman generation and four-wave
mixing~\cite{lettPRA08,phillipsPRA11}, we need to conduct the experiment
using a spatially configurable local oscillator and thus directly mapping
the output spatial mode of the squeezed vacuum.

In conclusion, we demonstrated that it is possible to generate a squeezed
vacuum with non-zero angular momentum via PSR squeezing by manipulating
the transverse profile of the pump beam before the vapor cell using a
phase mask.  We reported $1.7 \pm 0.2$~dB of squeezing in the first order
Laguerre-Gaussian spatial mode, which was comparable  to the  $1.8 \pm
0.2$~dB squeezing value observed in the same setup with a regular laser
pump field.  Thus, the change in the pump intensity distribution did not
change the maximum achieved value of squeezing, but only the experimental
conditions (atomic density and pump laser intensity) at which squeezing occurred,
so it might be possible to imprint spatial information into the squeezed
vacuum optical field by controlling the profile of the pump field using,
for example, a liquid crystal spatial light modulator.  We also
investigated the effect of self-defocusing that led to beam expansion
after interaction with Rb atoms at higher cell temperature.  While the
sizes of both Gaussian and Laguerre-Gaussian beams increased as the atomic
density increased, the overall shape was well preserved in the range of
explored experimental parameters.  In general, we found no clear
correlation between self-defocusing effect and generation or preservation
of squeezed states, although additional investigations were required.


The authors thank G.~A. Swartzlander for lending us the vortex phase mask,
and G. Romanov and T. Horrom for the assistance with the experiment.
This research was supported by AFOSR grant FA9550-13-1-0098.


%

\end{document}